\documentclass[twocolumn,aps,prl,showpacs,nobibnotes,
amsmath,amssymb]{revtex4-1}
\usepackage{natbib}
\usepackage{graphicx}
\usepackage{bm}
\usepackage{gensymb}
\usepackage{amsmath}
\usepackage{amssymb}
\usepackage{overpic}
\usepackage{color}
\usepackage{slashbox,multirow,booktabs}
\begin{document}
\title{Temperature dependence of long coherence times of oxide charge qubits}
%in oxide double quantum dots}
%Temperature dependence of long coherence times in  charge qubits}
%Temperature effects on decoherence in a two-site model for hard-core bosons}
\author{A. Dey}
%${^1}$}
\author {S. Yarlagadda}
%${^1}$}
\affiliation{
%${^1}$
CMP Div.,
1/AF Salt Lake, Saha Institute of Nuclear physics, Kolkata 700064, India}
%\affiliation{${^2}$Physical Sciences and Engineering Division, Argonne National Laboratory,
%Argonne, Illinois 60439, USA}
%03.67.Pp, 75.10.Jm, 87.10.Hk}}
\date{\today}
\begin{abstract}
The ability to maintain coherence and control in a qubit is a major requirement 
%the heart of making a scalable 
for quantum computation.
We show theoretically that long coherence times can be achieved above boiling point of liquid
helium in  charge qubits of oxide double quantum dots. Detuning the dots to 
a fraction of the optical phonon energy, increasing the
electron-phonon coupling, reducing
the adiabaticity, or decreasing the temperature
enhances the coherence time. We consider a system that is initially
decoupled from the phonon bath in the polaronic frame of reference and  solve the
non-Markovian quantum master equation; we find that the system decoheres
after a long time,
%; on the other hand, the population difference does not decay since
despite the fact that no energy is exchanged with the bath.

%We deal with a two-site, one hard-core boson (HCB) system strongly coupled to a site-dependent local optical-phonon environment. 
%Our analysis covers different temperature regimes of the phonon bath and its effect on the coherent dynamics of the system. 
%Additionally, the two sites are detuned with respect to each other by 
%introducing site-dependent potentials. In strong coupling limit, we make our analysis in polaronic frame of reference 
%and use the non-Markovian quantum master equation technique to study the dynamics of coherence as well as of the excited-state population . 
%We show that, at zero temperature 
%the coherence of the system shows recoherence when detuning energy is zero; 
%whereas, at finite temperature, the coherence irreversibly decays with time due to the thermal fluctuations and such an irreversible 
%decay of coherence increases with increasing temperature. Moreover, at finite temperature, 
%recoherence can be regained by turning on the detuning to a value that does not match with the phonon excitations. 
%For  particular values of temperature, HCB hopping strength and detuning potential, decoherence slows down as the HCB-phonon coupling increases; 
%furthermore, for a particular coupling, decoherence is most pronounced when the detuning is at resonance with twice the polaronic energy. 

\end{abstract}
\maketitle
%\section {Introduction}
{\em Introduction.}\textemdash
Construction of scalable quantum computers
has motivated identification of coherent two-level solid-state systems.
A simple solid-state two-level system is the charge qubit.
Charge qubit holds promise for high-speed manipulation
due to strong coupling of the electron to electric field.
On the other hand, large decoherence times have not been achieved so far in the
semiconductor double quantum dot systems studied as charge qubits 
\cite{cq_hirayama,cq_fujisawa1,cq_petta,cq_fujisawa2,cq_ritchie1,cq_ritchie2,cq_gossard,cq_guo1,cq_guo2,cq_copper1,cq_copper2}.
%;consequently, gates operating faster than decohering time scales set by noise is a challenging problem.
Furthermore,  the quantum dots employed in the
decoherence studies had a typical diameter of $\sim 200$ nm 
with the corresponding electron temperatures being $\sim 100$ mK.
Thus concomitant realization of fast operation and large coherence
times in small solid state qubits around boiling point of liquid  nitrogen (or at least liquid helium),
although very useful for quantum computation,
has been elusive so far. Here, as compared to a semiconductor double quantum
dot (DQD), we demonstrate that an oxide (e.g., manganite) based DQD yields 
significantly higher decoherence times
at higher temperatures and in smaller sized systems.
%\textcolor{red}{Reference to theory papers of charge qubits and nonequil. dynamics}

Compared to semiconductors, oxides offer a vastly richer physics 
involving diverse magnetic, charge, and orbital correlations \cite{dagotto,tokura,ahn1,ahn2}.
Owing to significantly smaller extent of the wavefunction, oxides
can meet the miniaturization demands much better than semiconductors.
Low-dimensional oxides present new opportunities for
devices where electronic, lattice, and magnetic properties can be optimized by engineering
many-body interactions, fields, geometries, disorder, strain, etc.
In this work, we illustrate the device potential of low-dimensional oxides
through our analysis of an oxide DQD as a charge qubit.

Decoherence is one of the main obstacles in quantum information processing 
that degrades  the precious resource of quantum mechanical 
superpositions \cite{schloss,zurek}. Because of system-environment interactions, the quantum information
of the system leaks out to the large number of degrees 
of freedom of the environment. The temperature of the environment affects 
the reduced system dynamics and introduces additional relaxation 
channel for the system. 
%A zero temperature analysis is valid
%when $\hbar \omega \gg K_{B}T$ with $\omega$, $K_{B}$ and $T$ being the phonon 
%frequency, the Boltzmann constant and bath temperature, respectively. As this condition
%is not always guaranteed, 
For practical implementation, the finite temperature 
situation needs to be investigated thoroughly for cases
such as boiling points of liquid helium and liquid nitrogen and room temperature.
In this paper,
we show that long decoherence times 
can be achieved in oxide charge qubits at these elevated
(above dilution refrigerator) temperatures 
in contrast to semiconductor charge qubits.

%Finite temperature considerations are relevant with the excited bath 
%states having considerable probabilities. For the system coupled only to the vacuum of the bath, the excited state can relax down to the ground 
%state by emitting energy spontaneously; whereas, when it couples to the excited states of the bath, the bath can provide excitation to the 
%system leading it to jump from the ground state to the excited state \cite{wang}.  
%For a boson detector (BD) model, the temperature effect renormalizes the interaction strength between the incident particle and the detector and 
%decoherence occurs due to thermal fluctuations \cite{sachiko}. Finite-temperature decoherence effects on the Casimir force between two conducting 
%plates and a topological Aharonov-Bohm (AB) force between fluxons in a superconductor have been studied \cite{nussinov}. It has been shown that 
%the Casimir force loses the dependence on the angle between conduction paths with a decoherence time of $\frac{h}{K_{B}T}$; furthermore, the 
%thermal fluctuations can strongly suppress the AB force by randomizing the electron's phase over a time period {$\mathcal {O}$}($\frac{h}{K_{B}T}$). 
%In an exciton-phonon system, the thermally-induced decoherence is enhanced with temperature and the recurrence of coherence gets diminished 
%as temperature increases; moreover, it has been shown that, a confined lattice favours low decoherence and recurrence compared to that in a 
%lattice with larger length scale \cite{vincent}. 
%The temperature value determines the relative importance of
As temperature is varied, two qualitatively different mechanisms are relevant for transport
in a  system of  electrons strongly coupled to optical phonons, 
namely, the band-like motion and the random hopping of small polarons \cite{holstein}. 
%Actually, in a many-site system,
At higher temperatures, the overlap between the simple-harmonic-oscillator wave functions 
of host molecules
on neighboring sites decreases 
because higher eigenfunctions with
more nodes come into play; consequently,
the polaron bandwidth decreases.
At higher temperatures, 
 the random process dominates over the band motion;  
the crossover from band-like motion to hopping
conduction occurs when the uncertainty in energy
(produced by electron-phonon scattering)
 is comparable to half the bandwidth \cite{alexandrov,yarlagadda}.
%the coherent band-like motion is completely destroyed leaving only the random 
%hopping processes \cite{alexandrov}. 
Here, at various temperatures,
we investigate in detail how coherence in a single electron 
(tunneling between two dots) is effected
 due to strong interaction with optical phonons.
{\em DQD model with environment.}\textemdash
We consider a laterally coupled DQD system 
for our two-level qubit.
The charge in the DQD system is denoted $(N_1,N_2)$ with $N_1$
and $N_2$ being the number of electrons on  dots 1 and 2, respectively. 
The quantum dots are taken to be identical with the same
%each possessing a
charging energy $E_C = e^2/C$ where $e$ is the
charge of an electron and $C$ is the capacitance between the dot
and its surroundings. The capacitance $C$ can be conservatively approximated
%(up to small factor 3)
by the self-capacitance
$C_0 = 4 \epsilon_m \epsilon_0 D$ \cite {kouwenhoven}
which for a manganite
dot with dielectric constant $\epsilon_m = 10$ and diameter $D= 10$ nm
yields $E_C \sim 0.05$ eV. We analyze
 situations where the thermal energy $k_B T$ as well as the
 the  detuning $\Delta \varepsilon \equiv \varepsilon_1- \varepsilon_2$
 (between the lowest energy levels  in the two dots) 
% as well as the thermal energy $k_B T$
are both  smaller than $E_C$ so that the 
  dynamics of a single electron
 can be studied when $|N_1-N_2| =1$.
 %; the  detuning $\Delta\epsilion \equiv \epsilon_1- \epsilon_2$
 %between the lowest energy levels
 %in the two dots is also taken to be
 %smaller than $E_C$.
  Consequently, we define the relevant charge
 states as
$|10\rangle \equiv (N+1,N)$ and $|01\rangle \equiv (N,N+1)$.

The coupled dots are described by the following Hamiltonian of a single electron tunneling between them:
\begin{eqnarray}
\!\!\!\!\!\!\!\!
H_{\rm DQD} &=&\varepsilon_{1} m_{1}+\varepsilon_{2} m_{2}
 -\frac{J_{\perp}}{2}(c_{1}^{\dagger}c_{2}+c_{2}^{\dagger}c_{1})
+ J_{\parallel}m_{1}m_{2} ,
%\nonumber \\
% && +g\omega\sum_{i=1,2} (n_{i}-\frac{1}{2})( a_i + a_{i}^{\dagger})
% + \omega \sum_{i=1,2} a_{i}^{\dagger} a_i,
\label{ham_dqd}
 \end{eqnarray}
 where the electron destruction operator in dot $i$ is defined as $c_i$ 
and $m_i\equiv c^{\dagger}_i c_i$. Furthermore,
 the energies $\varepsilon_i$ and the interdot tunnel coupling
 $\frac{J_{\perp}}{2}$ are adjusted by external gates; the nearest
 neighbor repulsion $J_{\parallel}$ is due to Coulomb interaction.
 %electron-electron and electron-phonon  interactions. 
The total Hamiltonian is expressed as
$H = H_{\rm DQD}+H_{\rm P}+H_{\rm EP}$ where
the additional term
$H_{\rm P} = \sum_{i,k} \omega_{k} a_{i,k}^{\dagger} a_{i,k} $ is due to the
optical phonon environment
while $H_{\rm EP} = \frac{1}{\sqrt{N}}\sum_{i,k}g_{k}\omega_{k} (m_{i}-\frac{1}{2})( a_{i,k} + a_{i,k}^{\dagger})$
is due to the electron-phonon interaction; here,
$a_{j,k}$ is the destruction operator of  mode k phonons at site j,
$g_{k}$ is the electron-phonon coupling strength,
and $\omega_k$ is the optical phonon frequency 
%\textcolor{red}
{with weak dispersion}.

In the strong coupling regime, to perform perturbation theory effectively, we locally displace
the harmonic oscillators by Lang-Firsov (LF) transformation  \cite{lf}
$H^{L}\equiv e^{S}He^{-S}$ with 
$S= -\frac{1}{\sqrt{N}}\sum_{i,k}g_{k} (m_i-\frac{1}{2}) (a_{i,k} - a_{i,k}^{\dagger})$.
In the LF frame, the electron is clothed with phonons reducing the tunnelling term $J_\perp$
in Eq. (\ref{ham_dqd}) 
 to $J^{\rm mf}_\perp\equiv J_\perp e^{-\frac{1}{{N}}
 \sum_{k}g_k^2 {\rm coth \frac{\beta\omega_k}{2}}}$.
This reduction of the polaronic
tunneling at enhanced temperatures occurs for the same reason as that in a polaron band.
 Therefore, in the DQD, the single particle energy is much smaller than the charging energy
 $E_C$.
 %{\textcolor{red}{(Check)}}.
 The redefined polaronic system, the bath environment with displaced harmonic oscillators,
 and the interaction term in
 the LF frame are respectively given by
 \begin{eqnarray}
H_s^{L}&=&\varepsilon_{1} m_{1}+\varepsilon_{2} m_{2}
-  \frac{J_\perp^{\rm mf}}{2} 
(c_{1}^{\dagger}c_{2}+c_{2}^{\dagger}c_{1})
+ J_{\parallel}m_{1}m_{2},\\
H_{R}^{L} &=&  \sum_{i,k} \omega_{k} a_{i,k}^{\dagger} a_{i,k} ,\\
{\rm and} \nonumber \\
 H_I^{L}&=& -\frac{1}{2} [J_\perp ^+ c_{1}^{\dagger}c_{2} + J_\perp ^- c_{2}^{\dagger}c_{1} ] ,
\label{hil1}
\end{eqnarray}
where the fluctuation of the local phonons around the mean phonon field $J_\perp ^{\rm mf}$
is given by $J_\perp ^{\pm} = J_\perp e^{\pm \frac{1}{\sqrt{N}}\sum_{k}~ g_k [(a_{2,k} - a_{2,k}^{\dagger})
-(a_{1,k} - a_{1,k}^{\dagger})]} -J_\perp ^{\rm mf}$.
%; here the term $J_\perp ^{\rm mf}$ is the mean phonon field at finite temperature. 
Since the small parameter is inversely proportional to the coupling strength \cite{hcbsite},
$H^{L}_{I}$ 
%\textcolor{red}
{is weak in the LF frame which suits a  perturbative treatment}.
%Furthermore, one can see that the polaronic
%tunneling is reduced at enhanced temperatures for the same reason as that in a polaron band.

%Actually, in a many-sitetem,
%the polaron bandwidth decreases 
%with increasing temperature as the overlap between the simple-harmonic-oscillator wave 
%functions of neighboring sites decreases with temperature; 
%at higher temperatures, the excited oscillator states appear with more number 
%of nodes and thus, adjacent wavefunctions destructively interfere \cite{dalibor,ys}. 

{\em Polaron dynamics.}\textemdash
The dynamics of the system is described in terms of the reduced density matrix
of the system $\rho_s(t) \equiv {\rm Tr}_R [\rho_T (t)]$
%$ at time $t$
%Tr B [ρ T (t)]
where the  degrees of freedom of the  bath are traced
out from the total system-environment density matrix $\rho_T (t)$.
We start with the simply separable initial
state $\rho_T(0) = \rho_s(0) \otimes R_0$ in the polaronic frame of reference
with the expectation that perturbation at large coupling will not produce much
change to the state of the system \cite{hcbsite}. Here,
$R_0$  is the phonon density matrix at thermal equilibrium given by 
$R_0=\sum_{\{n_{k}\}}|\{n_k\} \rangle_{ph}~\!_{ph}\langle \{n_{k}\}|e^{-\beta \bar{\omega}_{n}}/Z$; 
the phonon eigenstate and eigenenergy 
are given by $|\{n_k\} \rangle_{ph}\equiv |\{n^k_1\},\{n^k_2\} \rangle_{ph}$ 
and $\bar{\omega}_n\equiv \sum_k\omega_k(n^k_1+n^k_2)$ with $n^k_1$ and $n^k_2$ 
being the  mode k phonon occupation numbers 
in dots 1 and 2. This separable initial state can be obtained in a physical 
system such as an oxide-based DQD by using a small value of 
$J_{\perp}/\omega_k$ \cite{poldyn}.

We analyze the reduced dynamics of the system by the second-order, time-convolutionless,
%(TCL), 
non-Markovian, quantum-master equation in the interaction picture
[i.e., Redfield equation (see Ref. \onlinecite{Pet})]:
\begin{eqnarray}
\!\!\!\!\!\! {\frac{d \tilde{\rho}_s(t)}{dt} 
= - \int_0^t d\tau Tr_R[\tilde{H}_I^L(t),[\tilde{H}_I^L(\tau), \tilde{\rho}_s(t) \otimes R_0]].}
\label{mas_therm}
\end{eqnarray}
Here, an operator $A$ is expressed in the interaction picture representation
as $\tilde{A}(t) = e^{i(H_s^L+H^L_{R})t} A  e^{-i(H_s^L+H^L_{R})t}$.
For our analysis we use the eigenstate basis
%, i.e., 
\Big\{$|\epsilon_s\rangle=\frac{|10\rangle-|01\rangle}{\sqrt{2}}$, 
$|\epsilon_t\rangle=\frac{|10\rangle+|01\rangle}{\sqrt{2}}$\Big\} 
(with eigenenergies $\epsilon_s$ and $\epsilon_t$)
for zero detuning ($\Delta\varepsilon=0$) 
and the basis \Big\{$|10\rangle$, $|01\rangle$\Big\} for strong detuning 
($\Delta\varepsilon>>J_{\perp}^{\rm mf}$).
%and $g^2>>1$ are considered. 
For the  zero (finite) detuning case, to analyze coherence and population,
we solve for the offdiagonal density matrix 
element 
$\tilde{c}_{st}(t)\equiv\langle \epsilon_s|\tilde{\rho}_s(t)|\epsilon_t\rangle$
$\big (\tilde{c}_{10}(t)\equiv\langle 10|\tilde{\rho}_s(t)|01\rangle \big )$ 
 and 
the diagonal element 
$\tilde{p}_s(t)\equiv\langle \epsilon_s|\tilde{\rho}_s(t)|\epsilon_s\rangle$
$\big ( \tilde{p}_{10}(t)\equiv\langle 10|\tilde{\rho}_s(t)|10\rangle \big )$. 
%($\tilde{p}_s(t)\langle \epsilon_s|\tilde{\rho}_s(t)|\epsilon_s\rangle$).

\textit{Zero detuning.}\textemdash 
For the case when
%Here, we deal with the the case when 
$\Delta \varepsilon=0$,
%i.e., when energetically both the sites are same. 
from Eq. (\ref{mas_therm}), we get the following equations of motion 
for the offdiagonal and diagonal elements of $\tilde{\rho}_s (t)$.
%(see the Supplemental Material \cite{suppl}).
%\begin{widetext}
\begin{eqnarray}
&& \!\!\!\! \dot{\tilde{c}}_{st}(t) \nonumber \\
&& \!\!\!\! = -\Bigg[\tilde{c}_{st}(t)
% \sum_{\{n^k\},\{m^k\}; {\rm even},0}
 \sum_{\bar{n}+\bar{m}= {\rm even},0}
 \mathcal{J}_{nm}f_{nm}(t) \nonumber \\
 &&~~~-\frac{ie^{i\delta \epsilon t} }{4}\bigg(\tilde{c}_{st}(t) e^{-i\delta \epsilon t} \!\!\!\!
 %\sum_{\{n^k\},\{m^k\}; {\rm odd}}
 \sum_{\bar{n}+\bar{m}={\rm odd}}
\mathcal{J}_{nm} \Big \{\mathbb{F}^{+}_{nm}(t)-\mathbb{F}^{-*}_{nm}(t)\Big\}\nonumber \\ 
&&~~~~~~~~~~~~~~~~~~~~~~~~~~~~~~~~~~~~~~~~~~~~~~~~~~~~~ -
%e^{2i\delta \epsilon t} \times 
{\rm H.c.} \bigg )\Bigg], 
%\\
%&&~~~~~~+\tilde{c}^*(t)\frac{i}{4}\sum_{\{n^k\},\{m^k\}; {\rm odd}} \mathcal{J}_{nm} 
%e^{2i\delta \epsilon t} \Big \{\mathbb{{F}^{*+}}_{nm}(t)-\mathbb{{F}^{-}}_{nm}(t)
%\Big\}\Bigg]  ,\nonumber \\
\label{off_sing1}
\end{eqnarray}
and 
\begin{eqnarray}
&&\dot{\tilde{p}}_{s}(t) \nonumber \\&&=-\frac{1}{4i}
%\sum_{\{n^k\},\{m^k\};{\rm odd}}
\sum_{\bar{n}+\bar{m}={\rm odd}}
\mathcal{J}_{nm}
   \Bigg[ \tilde{p}_{s}(t)
    \bigg\{ \Big ( \mathbb{F}^{+}_{nm}(t)+\mathbb{F}^{-}_{nm}(t) \Big ) - {\rm H.c.} \bigg\} \nonumber \\
    %-\mathbb{F}^{*+}_{nm}(t)\Big\}+\nonumber \\ 
 %   &&~~~~~~~~~~~~~~\Big\{\mathbb{F}^{-}_{nm}(t)-\mathbb{F}^{*-}_{nm}(t)\Big\} \Big)
&&~~~~~~~~~~~~~~~~~~~~~~~~~~~~~~~~~~~~  - \Big\{\mathbb{F}^{-}_{nm}(t)- {\rm H.c.} \Big\} \Bigg],
    \label{dia_sing1} 
\end{eqnarray}
where, $\bar{n}\equiv \sum_k(n^k_1+n^k_2)$, $\bar{m}\equiv \sum_k(m^k_1+m^k_2)$,
$\mathcal{J}_{nm}=\Big ({_{ph}}\langle \{n_k\}|J^+_{\perp}|\{m_k\}\rangle_{ph}\Big )^2
\frac{e^{-\beta \bar{\omega}_{n}}}{Z}$, 
$f_{nm}(t)=\frac{\sin(\bar{\omega}_{n}-\bar{\omega}_m)t}{\bar{\omega}_{n}-\bar{\omega}_m}$, and 
$\mathbb{F}^{\pm}_{nm}(t)=\frac{e^{i(\bar{\omega}_{n}-
\bar{\omega}_m\pm\delta \epsilon)t}-1}{\bar{\omega}_{n}-\bar{\omega}_m\pm\delta \epsilon}$ 
%Here, 
%$\sum_{\{n^k\},\{m^k\}; {\rm even},0}$ ($\sum_{\{n^k\},\{m^k\}; {\rm odd}}$) 
%is the sum over the terms for which $|n^k_1+n^k_2-m^k_1-m^k_2|$ is even or zero (odd) 
%and
with $\delta\epsilon \equiv \epsilon_s-\epsilon_t$. 
%We numerically solve the differential equation for $c(t)$ and its complex 
%conjugate equation and plot the coherence factor 
%${\rm C_{st}(t)}\equiv c_{st}(t)/c_{st}(0)$ in Figs. \ref{fig1_thrm} (a) and \ref{fig4_thrm} (a).
%Furthermore, we also calculate population difference ${\rm P_{st}(t)}\equiv (2 p_{s}(t)-1)/(2 p_{s}(0) -1)$
% using Eq. (\ref{dia_sing1}) and compare its time evolution with ${\rm C_{st}(t)}$
%in Fig. \ref{fig2_thrm}.
To understand coherence and population evolution,
we define the coherence factor 
${\rm C_{st}(t)}\equiv c_{st}(t)/c_{st}(0)$ which can be obtained by solving Eq. (\ref{off_sing1})
and its complex conjugate; we also calculate the
population difference ${\rm P_{st}(t)}\equiv (2 p_{s}(t)-1)/(2 p_{s}(0) -1)$.

\textit{Finite detuning.}\textemdash 
As a strategy to mitigate decoherence, we employ sizeable energy detuning. 
For the case of finite detuning
%We consider the dots to be detuned with respect to each other with a potential
$\Delta\varepsilon >> \delta \epsilon$,
%much larger than $\delta \epsilon$. 
the equations for the offdiagonal and diagonal density matrix elements 
[obtained from Eq. (\ref{mas_therm})] are given by
 \begin{eqnarray}
 && \!\!\!\! \dot{\tilde{c}}_{10}(t)\nonumber \\
 && \!\!\!\! = -\frac{i}{4}\sum_{\{n^k\},\{m^k\}}\mathcal{J}_{nm} 
 \Bigg[\tilde{c}_{10}(t)\Big( \mathcal{F}^{-*}_{nm}(t)- \mathcal{F}^{+}_{nm}(t)\Big)\nonumber \\
 &&~~~~~~~~~~~~~~~~~~~~~~~~~+
 (-1)^{(\bar{n}+\bar{m})}
 %{\sum_k (n^k_1+n^k_2-m^k_1-m^k_2)}
 e^{2i\Delta\varepsilon t} \nonumber \\
 &&~~~~~~~~~~~~~~~~~~~~~~~~~~ \times
 \tilde{c}^{*}_{10}(t)\Big( \mathcal{F}^{-}_{nm}(t)- \mathcal{F}^{+*}_{nm}(t)\Big)
 %\times h.c.
 \Bigg] , 
 %\\
 %{\rm and} \nonumber \\
 %\\
 %&&~~~~~~~~~~~~~~~~~~~~~~~~~~~~~~~\times e^{2i\Delta\varepsilon t}
 %\Big(\mathcal{F}^{-}_{nm}(t)-\mathcal{F}^{*+}_{nm}(t) \Big)\Bigg]\\
 \label{off_therm1}
 \end{eqnarray}
 and 
 \begin{eqnarray}
&&\dot{\tilde{p}}_{10}(t) \nonumber \\&&=-\frac{1}{4 i}\sum_{\{n^k\},\{m^k\}}\mathcal{J}_{nm}
    \Bigg[ \tilde{p}_{10}(t)
    \bigg\{\Big ( \mathcal{F}^{+}_{nm}(t)+\mathcal{F}^{-}_{nm}(t) \Big ) -{\rm H.c.}\bigg\}\nonumber \\ 
%    &&~~~~~~~~~~~~~~\Big\{\mathcal{F}^{-}_{nm}(t)-\mathcal{F}^{*-}_{nm}(t)\Big\} \Big)
&&~~~~~~~~~~~~~~~~~~~~~~~~~~~~~~~~~  - \Big\{\mathcal{F}^{-}_{nm}(t)- {\rm H.c.} \Big\} \Bigg],
 \label{dia_therm1} 
\end{eqnarray}
where 
%$\mathcal{J}_{nm}=({_{ph}}\langle \{n_k\}|J^+_{\perp}|\{m_k\}\rangle_{ph})^2
%\frac{e^{-\beta \bar{\omega}_{n}}}{Z}$ and 
$\mathcal{F}^{\pm}_{nm}(t)=\frac{e^{i(\bar{\omega}_{n}-\bar{\omega}_m\pm\Delta \varepsilon)t}-1}
{\bar{\omega}_{n}-\bar{\omega}_m\pm\Delta \varepsilon}$.
To characterize the dynamics, we define the relevant coherence factor
${\rm C_{10}(t)}\equiv c_{10}(t)/c_{10}(0)$ and the population difference
${\rm P_{10}(t)}\equiv (2 p_{10}(t)-1)/(2 p_{10}(0) -1)$ which can be calculated from the
above two equations. Furthermore, when both $\Delta \varepsilon$ and $\delta \epsilon$
are non-negligible,
a general derivation of the matrix elements 
of the four terms on the right-hand side
of Eq. (\ref{mas_therm}) is given in  the Supplemental Material \cite{suppl}.

{\em Results and discussion.}\textemdash
In oxides such as the manganites, 
%(where the carriers are coupled predominantly only to optical phonons),
% the weak k-dependence of $g_k$ is quite valid. 
we approximate the density of states ${\rm D}(\omega_k)$
of our multimode baths by a generalization of the Einstein model
and take it to be a box function of small width 
$\omega_u -\omega_l$ ($= 0.1 \omega_u$) and height $\frac{1}{(\omega_u -\omega_l)}$
\begin{eqnarray}
  {\rm D}(\omega_k)g_k^2&=& g^2 \frac{N}{\omega_u - \omega_l} \Theta (\omega_k - \omega_l) \Theta(\omega_u - \omega_k) ,
 \label{density}
\end{eqnarray}
where $\Theta(\omega)$ is the unit step function. Notice that $\frac{1}{N}\sum_k g^2_k =g^2$.

In our calculations, we have employed experimentally realistic values of parameters in 
perovskite manganites.
For tunneling we chose  $J_{\perp}/\omega_u = 0.5~\&~2.8$
with phonon energy $\hbar \omega_u = 0.05 $ eV;
these values of $J_{\perp}$
%$J_{\perp} = 2.8 \omega_u$ is realized in manganites, $J_{\perp} = 0.5 \omega_u$
can be achieved by adjusting 
%tunneling using 
a gate voltage.
%As regards strong electron-phonon coulpings, we used $g=2.0~\&~2.5$.
%While $J_{\perp} = 2.8 \omega_u$ is realized in manganites, $J_{\perp} = 0.5 \omega_u$
%can be achieved by restricting tunneling using a gate voltage. 
There is compelling evidence of strong electron-phonon coupling
in  manganites \cite{lanzara,pbl1}.
As regards strong  coulpings, we used $g=2.0~\&~2.5$.
The strength of electron-phonon coupling can be varied by using different rare earth (RE)
elements (such as $La,~ Pr,~ \&~Nd$) in the oxide $RE_{1-x} Ca_xMnO_3$ \cite{pbl2}.
Thus, we can study decoherence for a reasonable range of small parameter values 
$J_{\perp}/(2 \sqrt{2}g\omega_u)$ \cite{hcbsite}.
Furthermore, using manganites around the ferromagnetic colossal magnetoresistive 
regime would aid controllability in the DQDs. 

It is important to note that, though an
exponential decay is the usual feature of unstable systems,
it can deviate for quantum systems at short times \cite{steve}. In fact,
similar to Ref. \cite{steve}, our  long-time behavior 
of coherence is also given  $\exp (-\alpha -t/\tau )$ where $\tau$ is the coherence time
and $\alpha >0$ is a constant much smaller than unity.

\begin{figure}
[t]
%\centerline{\includegraphics[angle=90,angle=90,angle=90,height=1.7in,width=3.0in]{offdiag_site.2_.8.eps}}
%\vspace{-1.0cm}\centerline{\includegraphics[angle=90,angle=90,angle=90,height=1.7in,width=3.0in]{offdiag_site0.eps}}
\centerline{\includegraphics[angle=90,angle=90,angle=90,
width=3.2in]{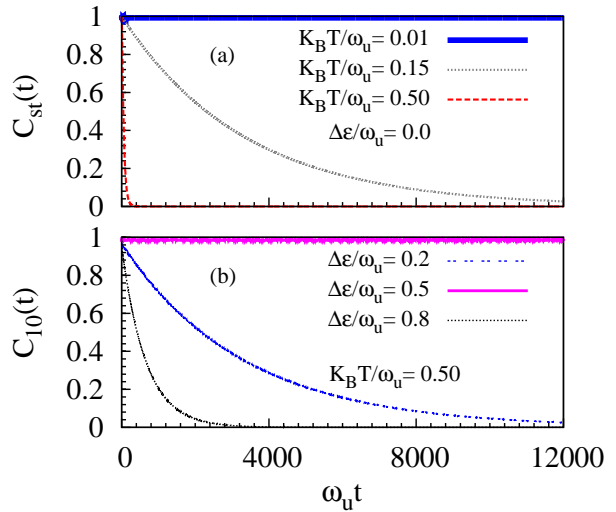}}
\caption{Exponential decay of coherence as a function of dimensionless time $\omega_ut$ plotted
for (a) detuning energy $\Delta \varepsilon=0.0$ (where largest decoherence is expected)
and at values
of dimensionless thermal energy ${\rm K_BT}/{\omega_u}= 0.01$ (near boiling point of liquid helium), $= 0.15$ 
(near boiling point of liquid nitrogen), and $=0.5$ (near room temperature);
for (b) dimensionless thermal energy ${\rm K_BT}/{\omega_u} = 0.5$ 
%(close to room temperature)
at various detuning energies $\Delta \varepsilon /\omega_u$
reflecting different 
%proximities to 
probabilities for resonance.
All the plots are at  values of adiabaticity
$J_{\perp}/{\omega_u} =0.5$ and electron-phonon coupling $g=2$ 
that are realizable in oxide DQDs.}
\label{fig1_thrm}
\end{figure}

\begin{figure}
[b]
\centerline{\includegraphics[angle=90,angle=90,angle=90,
width=3.3in
]{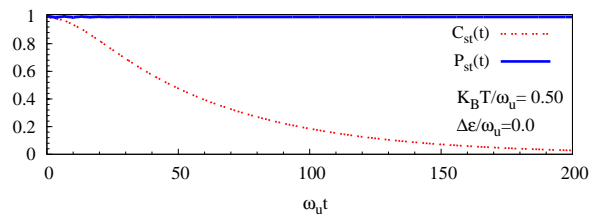}}
\caption{Time dependence, of coherence and population difference, depicting exponential decay of
coherence while population remains unchanged for the case of maximum decoherence 
(i.e., detuning $\Delta \varepsilon =0$) at around room temperature (i.e., ${\rm K_BT}/\omega_u =0.5$).
Adiabaticity
$J_{\perp}/{\omega_u} =0.5$ and electron-phonon coupling $g=2$ in the plot.}
\label{fig2_thrm}
\end{figure}

In Fig. \ref{fig1_thrm} (a) we investigate nature of decoherence for
zero detuning and various
%with varying
temperatures. At a very low  temperature 
${\rm K_BT}=0.01\omega_u$
(i.e., around boiling point of liquid
helium), we see that coherence does not decay; 
whereas, with increasing temperature it decays more rapidly. 
We compared numerical values of coherence 
for temperatures $0.01\omega_u$, with those at much lower temperatures including $0$ K. Over the
entire time range in Fig. \ref{fig1_thrm}, 
with respect to the zero temperature case, we find that the coherence values for
${\rm K_BT}=0.01\omega_u$ do not change at least up to the
twelfth decimal place. 
This leads us to infer that $\tau> 100$ s at these low temperatures (see Table. \ref{table}). 
Similarly, at finite detuning values as well, by contrasting coherence at  temperature $0.01\omega_u$
with those at much lower temperatures, we report large coherence time
$\tau> 100$ s in Table. \ref{table}.
With increasing temperatures, not only do excited phonon states appear
with enhanced thermal probability 
but also the number of degenerate phonon eigenstates increases.
Even if the total phonon bath does not exchange excitation with the system
(since $\delta \epsilon << \omega_u$),
this leads to a fluctuation in local phonon 
excitations causing destruction of coherence;
consequently, there is a decay in coherence while the population
difference remains unchanged as shown in Fig. \ref{fig2_thrm}.
%\textcolor{red}
{Since there is no exchange of energy between the bath 
and the polaron, other unoccupied single particle states will not be
relevant in producing decoherence}.
The term $f_{nm}(t)$ in Eq. (\ref{off_sing1}) represents 
the contributions from degenerate excited phonon eigenstates. 
%We observe that the term containing 
%$f_{nm}(t)$ in Eq. (\ref{off_sing1}) indicates this decoherence process. 
At temperatures ${\rm K_BT}\ll\omega_u$, 
the phonon ground state (although being probabilistically dominant)
produces no decoherence as the strength of decoherence 
$\mathcal{J}_{00}=0$; furthermore, the next dominant term is proportional
to $\sim \exp(-\beta\omega_u)$ becomes non-negligible only at much higher
temperatures (i.e., ${\rm K_BT}/\omega_u= 0.15 ~\& ~0.5$).

\begin{figure}
[t]
\centerline{\includegraphics[angle=90,angle=90,angle=90,width=3.0in]{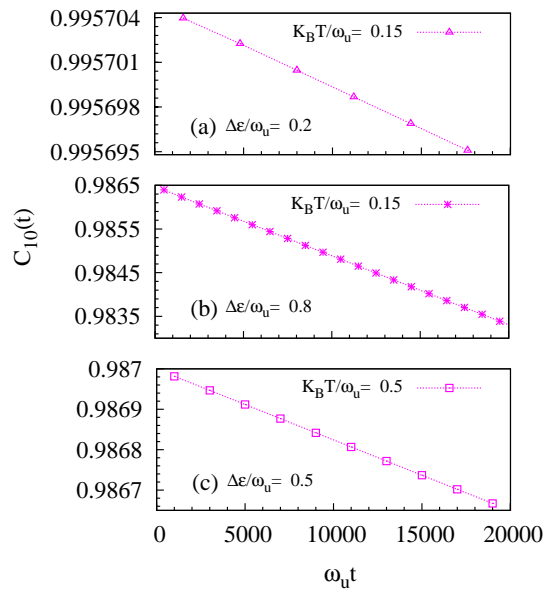}}
\caption{Depiction that  exponential decay of coherence is linear at small times.
The coherence time is inversely proportional to the slope.
The coherence values are averaged over intervals 
of width between successive points.
At a fixed temperature, among the detunings considered, the least decoherence  occurs 
when detuning $\Delta \varepsilon/\omega_u=0.5$
since it 
%is farthest from
has the lowest chance for resonance.
Figures were drawn at adiabaticity
$J_{\perp}/{\omega_u} =0.5$ and electron-phonon coupling $g=2$.}
\label{fig3_thrm}
\end{figure}

To obtain long coherence times even at elevated temperatures, 
we introduce detuning in the DQD and plot the coherence factor in Fig. \ref{fig1_thrm} (b)
at around the room temperature (i.e., ${\rm K_BT}=0.5\omega_u$). Here, we see that the decoherence 
time is much longer for $\Delta \varepsilon/\omega_u=0.5$ compared to the
other two cases $\Delta \varepsilon/\omega_u=0.2 ~\&~0.8$.
The phonon excitation $\bar{\omega}_n-\bar{\omega}_m=0.2\omega_u$ 
($\bar{\omega}_n-\bar{\omega}_m=0.8\omega_u$)
produces decoherence for $\Delta \varepsilon/\omega_u=0.2$ ($\Delta \varepsilon/\omega_u=0.8$)
case. Given the small phonon frequency window $\omega_u-\omega_l=0.1\omega_u$, 
as can be seen from Eq. (\ref{off_therm1}),
the thermal probability for the phonon excitation $0.8\omega_u$ is higher
[i.e., $\sim \exp(-\beta\omega_u)$] compared to that for $0.2\omega_u$
[i.e., $\sim \exp(-2\beta\omega_u)]$. 
On the other hand, decoherence time for $\Delta \varepsilon/\omega_u=0.5$ 
in Fig. \ref{fig1_thrm} (b) is the longest among all three
because the relevant thermal probability $\sim \exp(-4 \beta\omega_u)$ is comparatively smaller.

Next, we plot Fig. \ref{fig3_thrm}, by exploiting the linearity of the exponential
decay at times much smaller than the large decoherence times
realized near boiling point of liquid nitrogen 
(i.e., ${\rm K_BT}=0.15\omega_u$) at detuning $\Delta \varepsilon/\omega_u=0.2 ~\&~0.8$
and for  ${\rm K_BT}=0.5\omega_u$ at $\Delta \varepsilon/\omega_u=0.5$.
Fig. \ref{fig3_thrm}  is in 
%tune
agreement with Fig. \ref{fig1_thrm}. The numerical
values of coherence times for the cases in Figs. \ref{fig1_thrm} and \ref{fig3_thrm}
are reported in Table. \ref{table}; at temperatures much above boiling point of liquid helium,
we  see that with a properly chosen detuning one can get a 
%finite-temperature
decoherence 
time many orders of magnitude 
larger than the zero-detuning case.

\begin{figure}
%[b]
[h!]
%[!htb]
%\centerline{{\includegraphics[angle=90,angle=90,angle=90,height=3.in,width=1.5in]{offdiag_site0_temp.5.eps}}
%\centerline
{\includegraphics[angle=90,angle=90,angle=90,
width=3.3in]{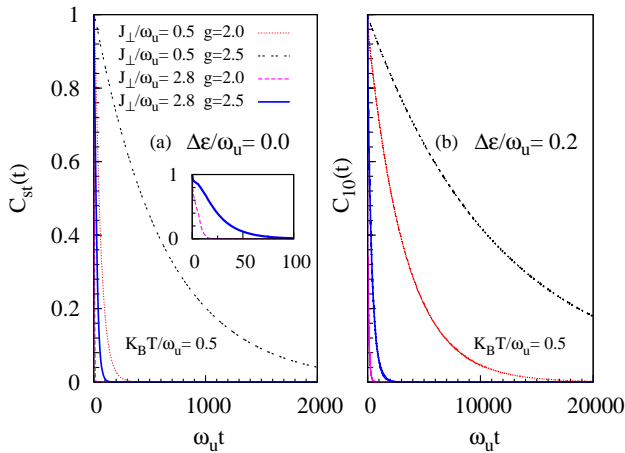}}
\caption{Plot showing  coherence decays faster (slower) with increasing adiabaticity $J_{\perp}/{\omega_u}$ 
(electron-phonon coupling $g$)
at two representative values of detuning, i.e., $\Delta \varepsilon/\omega_u=0.0~ \&~ 0.2$,
and near room temperature.}
\label{fig4_thrm}
\end{figure}

\begin{table}
[t]
%[!htb]
\begin{center}
{
\begin{tabular}{ccccc}
%\toprule
 \noalign{\global\arrayrulewidth.8pt}
\hline\hline
 \noalign{\global\arrayrulewidth .8pt}
\backslashbox
{\scalebox{1.2}{${{\rm K_{B}T}/ \omega_u}$}}{\vspace{-0.3cm}\scalebox{1.2}{$\Large {\Delta \varepsilon/\omega_u}$}}&~~ ${0.0}$&  
~~ ${0.2}$& ~~${0.5}$&~~${0.8}$ \\ 
\hline{} 
%\multicolumn{1}{c}{$g$}&
%\multicolumn {2}{c}{$0$} &
%\multicolumn {2}{c}{$\Delta \varepsilon/\omega=2.5$} &
%\multicolumn {2}{c}{$\Delta \varepsilon/\omega=2.9$}\\ [1.0ex] 
%\cline{1-7} 
%{} &{ ${\rm C_\infty}$}~~~ {${\rm P_\infty}$} &~~   { ${\rm C_\infty}$}~~~ {${\rm P_\infty}$}&~~   { ${\rm C_\infty}$}~~~ {${\rm P_\infty}$}\\
%\hline
\!\!\!\!\!\!\!\!\!\!\!\!\!\!\!\!\!\!\!\!\!\!\!\!\!\!\!\!$0.01$  &  $ >\!\!100$ s 
&~   $>\!\!100$ s &~  $>\!\!100$ s & $>\!\!100$ s \\
\!\!\!\!\!\!\!\!\!\!\!\!\!\!\!\!\!\!\!\!\!\!\!\!\!\!$0.15$&~~ 50 ps &~~~   24 $\mu$s &~  $>\! 0.1$ s &~ 83 ns \\
\!\!\!\!\!\!\!\!\!\!\!\!\!\!\!\!\!\!\!\!\!\!\!\!\!\!\!\! $0.50$& ~~~~1 ps &~~~ 47 ps &~\!  0.75 $\mu$s &~ 10 ps \\
\noalign{\global\arrayrulewidth.8pt}
\hline\hline
 \noalign{\global\arrayrulewidth .8pt}
\end{tabular}}
\caption{{Coherence times at various values of scaled thermal energy ${\rm K_BT}/\omega_u$ and detuning
energy $\Delta\varepsilon/\omega_u$  when  $\hbar \omega_u = 0.05$ eV.
}}
\label{table}
\end{center}
\end{table}

Lastly, in Fig. \ref{fig4_thrm}, we plot the coherence factor for different values of 
the coupling $g$
and the tunneling amplitude $J_{\perp}$. For a fixed tunneling, coherence is maintained for a longer time when the 
electron-phonon coupling is stronger. On the other hand, at a fixed value of the coupling, decoherence is
enhanced when tunneling increases. 
These results are consistent with the fact that decoherence diminishes at lower
values of the small 
parameter $\frac{J_{\perp}}{2\sqrt{2}g\omega}$ \cite{poldyn,hcbsite}.
Comparing Figs. \ref{fig4_thrm} (a) and (b), 
we again see that finite-detuning provides longer coherence times.

{\em Conclusion.}\textemdash
%Our study would be helpful in understanding and controlling decoherence in physical 
%systems such as  manganite DQDs in different temperature regimes (i.e., near liquid helium, liquid nitrogen,
%and  room temperatures).  
%It is important to note that, though an
%exponential decay is the usual feature of unstable systems,
%it can deviate for quantum systems at short times \cite{steve}. In fact,
%similar to Ref. \cite{steve}, our  long-time behavior 
%of coherence is also given  $\exp (-\alpha -t/\tau )$ where $\tau$ is the coherence time
%and $\alpha >0$ is a constant much smaller than unity.
% For an optical phonon environment, our work demonstrates that
%detuning energy  (when it is a sizeable fraction of a single phonon energy) actually
%significantly enhances the coherence time compared to the zero detuning case.
%This is in contrast to the observation in semiconductor charge DQDs  
%that coherence time decreases with increasing detuning \cite{cq_hirayama,cq_copper2}. 
%Next, although 
Although dynamics of a carrier coupled to optical phonons 
in a Holstein model has
been studied recently at both weak and strong couplings \cite{vidmar,eckstein}, nevertheless, it has been
done for the initial condition where the particle is uncoupled to the phonons
in the laboratory frame; furthermore, similar studies need to
be conducted for oxide DQD systems.

Our analysis shows that oxides (such as manganites) provide a useful material platform for realizing
charge qubits with long coherence times at elevated temperatures (i.e., higher than boiling point of liquid
helium). However, experimental confirmation is needed to clearly establish that
our oxide variant of double quantum dot  has an applicable
combination of maneuverability and coherence time; for universal quantum computation,
 it  needs to be demonstrated that high-fidelity gate operations, including
 two-qubit gate operations, can be performed.  
 
 We thank P. B. Littlewood, R. Ramesh. T. V. Ramakrishnan, A. J. Leggett, A. Bhattacharya, M. Manfra, J. Levy, and J. N. Eckstein
for stimulating discussions.

\clearpage
\newpage
\def \s{\sigma}
\def \sb{\bar\sigma}
\def \be{\begin{eqnarray}}
\def \ee{\end{eqnarray}}
\def \del{\partial}
\def \d{\dagger}
\def \sd{\sum\limits}
\def \sp{\!\!\!}
\def \spb{\sp\sp}
\setcounter{equation}{0}
\setcounter{figure}{0}
\renewcommand{\theequation}{S\arabic{equation}}
\renewcommand{\thefigure}{S\arabic{figure}}
%\begin{document}
\begin{widetext}
\noindent {{\bf Supplemental Material for \\
\\ 
\indent ``Temperature dependence of long coherence times of oxide charge qubits''}\\
\\
\indent A. Dey and S. Yarlagadda}
%\maketitle
\section{Details of the calculation of matrix elements}

Here we show the detailed calculations for the matrix elements obtained from the quantum master equation 
given by Eq. (5) in the main text. 
The double commutator in Eq. (5) can be broken into four terms.
In the equation for the matrix element $\langle 10| \tilde{\rho}_s(t)|01\rangle$,
the first term on the right-hand side [based on Eq. (5)] is given by
\begin{eqnarray}
 \langle 10|\sum_{n}~_{ph}\langle n |
 \tilde{H}_I^L(t)\tilde{H}_I^L(\tau)| n \rangle_{ph} \tilde{\rho}_s(t) \frac{e^{-\beta \bar{\omega}_{n}}}{Z}|01\rangle
 &=&\Big[\xi_{+, +}(\mathcal{T})~ \langle 10|\tilde{c}^+_{1}(t)\tilde{c}^-_{2}(t)
 \tilde{c}^+_{1}(\tau)\tilde{c}^-_{2}(\tau)\tilde{\rho}_s(t)|01\rangle \nonumber \\
 &&~~~+\xi_{+, -}(\mathcal{T})~ \langle 10|\tilde{c}^+_{1}(t)\tilde{c}^-_{2}(t)
 \tilde{c}^+_{2}(\tau)\tilde{c}^-_{1}(\tau)\tilde{\rho}_s(t)|01\rangle \nonumber \\
 &&~~~+\xi_{-, +}(\mathcal{T})~\langle 10|\tilde{c}^+_{2}(t)\tilde{c}^-_{1}(t)
 \tilde{c}^+_{1}(\tau)\tilde{c}^-_{2}(\tau)\tilde{\rho}_s(t)|01\rangle \nonumber \\
 &&~~~+\xi_{-, -}(\mathcal{T})~ \langle 10|\tilde{c}^+_{2}(t)\tilde{c}^-_{1}(t)
 \tilde{c}^+_{2}(\tau)\tilde{c}^-_{1}(\tau)\tilde{\rho}_s(t)|01\rangle\Big]. \nonumber \\
 \label{1st_mntxt_s}
 \end{eqnarray}
% \textcolor{red}
{Here, $\mathcal{T}=t-\tau$, i.e., the difference of times at which the the two-time correlation functions 
 $\xi_{\mu, \nu}(\mathcal{T})$ are calculated.} 
 The correlation functions can be written as
\begin{eqnarray}
 &&\xi_{\mu,\nu}(\mathcal{T})\nonumber \\
 &&=\frac{1}{4}\sum_{\{n_k\}}~ _{ph}\langle \{n_k\} |\tilde{J}^{\mu}_{\perp}
 (\mathcal{T})\tilde{J}^{\nu}_{\perp}| \{n_k\} \rangle_{ph}  \frac{e^{-\beta \bar{\omega}_{n}}}{Z} \nonumber \\
 &&=\frac{1}{4}\sum_{\{n_k\},\{m_k\}}~ _{ph}\langle \{n_k\} |\tilde{J}^{\mu}_{\perp}
 (\mathcal{T})|\{m_k\}\rangle_{ph}~_{ph}\langle \{m_k\} |\tilde{J}^{\nu}_{\perp}| \{n_k\} \rangle_{ph}  \frac{e^{-\beta \bar{\omega}_{n}}}{Z}\nonumber \\
 &&=\frac{1}{4}\sum_{\{n_k\},\{m_k\}}~ _{ph}\langle \{n_k\} |{J}^{\mu}_{\perp}
 |\{m_k\}\rangle_{ph}~_{ph}\langle \{m_k\} |{J}^{\nu}_{\perp}| \{n_k\} \rangle_{ph} e^{i(\bar{\omega}_n-\bar{\omega}_m)\mathcal{T}} 
 \frac{e^{-\beta \bar{\omega}_{n}}}{Z},
\end{eqnarray}
where we observe that
\begin{eqnarray}
 &&{_{ph}}\langle m_k|J^-_{\perp}|n_k\rangle_{ph}={_{ph}}\langle n_k|J^+_{\perp}|m_k\rangle_{ph}, \nonumber \\
 &&{\rm and} \nonumber\\
 &&{_{ph}}\langle m_k|J^+_{\perp}|n_k\rangle_{ph}=(-1)^{(n^k_1+n^k_2-m^k_1-m^k_2)}{_{ph}}\langle n_k|J^+_{\perp}|m_k\rangle_{ph}.\nonumber \\
 \label{matcon}
\end{eqnarray}
 Now, we calculate the phonon correlation function $\xi_{+, +}(\mathcal{T})$ below:
 \begin{eqnarray}
  && 
  \!\!\!\!\!\!\!\!
  \xi_{+, +}(\mathcal{T}) \nonumber \\
  &&
   \!\!\!\!\!\!\!\!
  = \frac{J^2_{\perp}}{4}\Bigg[ \sum_{\{n_k\}}~ _{ph}\langle \{n_k\} | e^{-\frac{1}{\sqrt{N}}\sum_{k}g_{k}[\{a_{1,k}(\mathcal{T})-a^{\dagger}_{1,k}(\mathcal{T})\}
  -\{a_{2,k}(\mathcal{T})-a^{\dagger}_{2,k}(\mathcal{T})\}]}\times e^{-\frac{1}{\sqrt{N}}\sum_{k}g_k[\{a_{1,k}-a^{\dagger}_{1,k}\}
  -\{a_{2,k}-a^{\dagger}_{2,k}\}]}|\{n_k\}\rangle_{ph} 
  %\nonumber \\
  %&&~~~~~~~~~~~~~~~~~~~~~~~~~~~~~~~~~~~~~~~~~~~~~~~~~~~~~~~~~~~~~~~~~~~~~~~~~~~~~~~~~~~~~~~~~~~~~~~~~~~~~~~~~~~~~~~~~~~~\times
  \frac{ e^{-\beta \bar{\omega}_{n}}}{Z} \nonumber \\
 &&~~~~~~~~~ -e^{-\frac{1}{{N}}\sum_{k}g_k^2 {\rm coth}(\frac{\beta\omega_{k}}{2})}\sum_{\{n_k\}}~ _{ph}\langle \{n_k\} |e^{-\frac{1}{\sqrt{N}}\sum_{k}g_{k}[\{a_{1,k}-a^{\dagger}_{1,k}\}
  -\{a_{2,k}-a^{\dagger}_{2,k}\}]}|\{n_k\}\rangle_{ph} \frac{ e^{-\beta \bar{\omega}_{n}}}{Z} \nonumber \\
 &&~~~~~~~~~ -e^{-\frac{1}{{N}}\sum_{k}g_k^2 {\rm coth}(\frac{\beta\omega_k}{2})}\sum_{\{n_k\}}~ _{ph}\langle \{n_k\} |e^{-\frac{1}{\sqrt{N}}\sum_{k}g_k[\{a_{1,k}(\mathcal{T})-a^{\dagger}_{1,k}(\mathcal{T})\}
  -\{a_{2,k}(\mathcal{T})-a^{\dagger}_{2,k}(\mathcal{T})\}]}|\{n_k\}\rangle_{ph} \frac{ e^{-\beta \bar{\omega}_{n}}}{Z} \nonumber \\
 &&~~~~~~~~~+ e^{-\frac{2}{{N}}\sum_{k}g_k^2 {\rm coth}(\frac{\beta\omega_{k}}{2})} \sum_{\{n_k\}} \frac{e^{-\beta\bar{\omega}_n}}{Z}\Bigg]
 \label{corr1}
 \end{eqnarray}
The first term in Eq. (\ref{corr1}) is written as
\begin{eqnarray}
 &&\frac{J^2_{\perp}}{4}\sum_{\{n_k\}}~ _{ph}\langle \{n_k\} | e^{-\frac{1}{\sqrt{N}}\sum_{k}g_k[\{a_{1,k}(\mathcal{T})-a^{\dagger}_{1,k}(\mathcal{T})\}
  -\{a_{2,k}(\mathcal{T})-a^{\dagger}_{2,k}(\mathcal{T})\}]}\times e^{-\frac{1}{\sqrt{N}}\sum_{k}g_k[\{a_{1,k}-a^{\dagger}_{1,k}\}
  -\{a_{2,k}-a^{\dagger}_{2,k}\}]}|\{n_{k}\}\rangle_{ph} \frac{ e^{-\beta \bar{\omega}_{n}}}{Z} \nonumber \\
  &&=\frac{J^2_{\perp}}{4}\sum_{\{n^k_1\}}~ _{ph}\langle \{n^k_1\} | e^{-\frac{1}{\sqrt{N}}\sum_{k}g_k\{a_{1,k}(\mathcal{T})-a^{\dagger}_{1,k}(\mathcal{T})\}}
  e^{-\frac{1}{\sqrt{N}}\sum_{k}g_k\{a_{1,k}-a^{\dagger}_{1,k}\}}|\{n^k_1\}\rangle_{ph} \frac{ e^{-\beta \bar{\omega}_{n_1}}}{Z_1}  \nonumber \\
 &&~~~~~~~ \times \sum_{\{n^k_2\}}~ _{ph}\langle \{n^k_2\} | e^{-\frac{1}{\sqrt{N}}\sum_{k}g_k\{a_{2,k}(\mathcal{T})-a^{\dagger}_{2,k}(\mathcal{T})\}}
  e^{-\frac{1}{\sqrt{N}}\sum_{k}g_k\{a_{2,k}-a^{\dagger}_{2,k}\}}|\{n^k_2\}\rangle_{ph} \frac{ e^{-\beta \bar{\omega}_{n_2}}}{Z_2} \nonumber \\
  &&=\frac{J^2_{\perp}}{4} e^{-\frac{2}{{N}}\sum_{k}g_k^2 {\rm coth}(\frac{\beta\omega_k}{2})[1+{\rm cos}(\omega_{k} \mathcal {T})]} 
  e^{\frac{2}{{N}}i\sum_{k}g_k^2 {\rm sin}(\omega_k \mathcal{T})},
  \label{1st}
\end{eqnarray}
where $Z_{j}=\sum_{\{n^k_j\}} e^{-\beta \bar{\omega}_{n_j}}$ and 
\begin{eqnarray}
 &&\sum_{\{n^k_1\}}~ _{ph}\langle \{n^k_1\} | e^{-\frac{1}{\sqrt{N}}\sum_{k}g_k\{a_{1,k}(\mathcal{T})-a^{\dagger}_{1,k}(\mathcal{T})\}}
  e^{-\frac{1}{\sqrt{N}}\sum_{k}g_k\{a_{1,k}-a^{\dagger}_{1,k}\}}|\{n^k_1\}\rangle_{ph} \frac{ e^{-\beta \bar{\omega}_{n_1}}}{Z_1} \nonumber \\ &&=
 e^{-\frac{1}{{N}}\sum_{k}g_k^2}\sum_{\{n^k_1\}}~ _{ph}\langle \{n^k_1\} | e^{\frac{1}{\sqrt{N}}\sum_{k}g_ka^{\dagger}_{1,k} e^{i\omega_k \mathcal{T}}}  e^{-\frac{1}{\sqrt{N}}\sum_{k}g_ka_{1,k} e^{-i\omega_k \mathcal{T}}} e^{\frac{1}{\sqrt{N}}\sum_{k}g_k a^{\dagger}_{1,k}} 
 e^{-\frac{1}{\sqrt{N}}\sum_{k}g_k a_{1,k}} | \{n^k_1\}\rangle_{ph} \frac{ e^{-\beta \bar{\omega}_{n_1}}}{Z_1} \nonumber \\
 &&= e^{-\frac{1}{{N}}\sum_k g_k^2(1+e^{i\omega_k \mathcal{T}})}\sum_{\{n^k_1\}}~ _{ph}\langle \{n^k_1\} | e^{\frac{1}{\sqrt{N}}\sum_k g_k a^{\dagger}_{1,k} (1+e^{i\omega_k \mathcal{T}})} e^{-\frac{1}{\sqrt{N}}\sum_{k} g_k a_{1,k} 
 (1+e^{i\omega_k \mathcal{T}})} |\{n^k_1\}\rangle_{ph}  \frac{ e^{-\beta \bar{\omega}_{n_1}}}{Z_1} \nonumber \\
 &&=e^{-\frac{1}{{N}}\sum_{k} g_k^2{\rm coth}(\frac{\beta\omega_k}{2})(1+{\rm cos}\omega_{k} \mathcal{T})}e^{i\frac{1}{{N}}\sum_{k} g_k^2{\rm sin}\omega_k\mathcal{T}}.
\end{eqnarray}

The second and third terms in Eq. (\ref{corr1}) are written as
\begin{eqnarray}
&&e^{-\frac{1}{{N}}\sum_{k}g_k^2 {\rm coth}(\frac{\beta\omega_k}{2})}\sum_{\{n_k\}}~ _{ph}\langle \{n_k\} |e^{-\frac{1}{\sqrt{N}}\sum_{k}g_k[\{a_{1,k}-a^{\dagger}_{1,k}\}
  -\{a_{2,k}-a^{\dagger}_{2,k}\}]}|\{n_k\}\rangle_{ph} \frac{ e^{-\beta \bar{\omega}_{n}}}{Z}\nonumber \\
&& = e^{-\frac{1}{{N}}\sum_{k}g_k^2 {\rm coth}(\frac{\beta\omega_k}{2})}\sum_{\{n_k\}}~ _{ph}\langle \{n\} |e^{-\frac{1}{\sqrt{N}}\sum_{k} g_k[\{a_{1,k}(\mathcal{T})-a^{\dagger}_{1,k}(\mathcal{T})\}
  -\{a_{2,k}(\mathcal{T})-a^{\dagger}_{2,k}(\mathcal{T})\}]}|\{n_k\}\rangle_{ph} \frac{ e^{-\beta \bar{\omega}_{n}}}{Z} \nonumber \\
 &&= \frac{J^2_{\perp}}{4} e^{-\frac{2}{{N}}\sum_{k} g_k^2 {\rm coth}(\frac{\beta\omega_k}{2})}.
 \label{2nd_3rd}
\end{eqnarray}
The fourth term in Eq. (\ref{corr1}) is 
\begin{eqnarray}
 \frac{J^2_{\perp}}{4} e^{-\frac{2}{{N}}\sum_{k}g_k^2 {\rm coth}(\frac{\beta\omega_k}{2})}\sum_{\{n_k\}} \frac{ e^{-\beta \bar{\omega}_{n}}}{Z}
 =\frac{J^2_{\perp}}{4} e^{-\frac{2}{{N}}\sum_{k}g_k^2 {\rm coth}(\frac{\beta\omega_k}{2})}.
 \label{4th}
\end{eqnarray}
Finally we get the simplified expression for $\xi_{+, +}(\mathcal{T})$
\begin{eqnarray}
  \xi_{+, +}(\mathcal{T})={\kappa}^2 
 \Big[e^{-\frac{2}{{N}}\sum_{k}g_k^2 {\rm coth}(\frac{\beta\omega_k}{2}){\rm cos}(\omega_k \mathcal {T})} 
  e^{\frac{2}{{N}}i\sum_k g_k^2 {\rm sin}(\omega_k \mathcal{T})}-1\Big],
  \label{corr1_final}
\end{eqnarray}
where $\kappa=\frac{J^{\rm mf}_{\perp}}{2}$.
%\begin{eqnarray}
 %\xi_{+, +}(\mathcal{T})=\kappa^2 
% \Big[e^{-2g^2 {\rm coth}(\frac{\beta\omega}{2}){\rm cos}(\omega \mathcal {T})} 
 % e^{2ig^2 {\rm sin}(\omega \mathcal{T})}-1\Big].
 % \label{corr1_final}
%\end{eqnarray}
In a similar fashion, the other correlation function can be written as
\begin{eqnarray}
  \xi_{+, -}(\mathcal{T})={\kappa}^2 
 \Big[e^{\frac{2}{{N}}\sum_{k}g_k^2 {\rm coth}(\frac{\beta\omega_k}{2}){\rm cos}(\omega_k \mathcal {T})} 
  e^{-\frac{2}{{N}}i\sum_k g_k^2 {\rm sin}(\omega_k \mathcal{T})}-1\Big].
  \label{corr2_final}
\end{eqnarray}
%\begin{eqnarray}
 %\xi_{+, -}(\mathcal{T})=\kappa^2 
 %\Big[e^{2g^2 {\rm coth}(\frac{\beta\omega}{2}){\rm cos}(\omega \mathcal {T})} 
 % e^{-2ig^2 {\rm sin}(\omega \mathcal{T})}-1\Big].
%  \label{corr2_final}
%\end{eqnarray}
Using Eqs. (\ref{corr1_final}) and (\ref{corr2_final}) we write Eq. (\ref{1st_mntxt_s}) as
\begin{eqnarray}
  &&\langle 10|\sum_{n}~_{ph}\langle n |
 \tilde{H}_I^L(t)\tilde{H}_I^L(\tau)| n \rangle_{ph} \tilde{\rho}_s(t) \frac{e^{-\beta \bar{\omega}_{n}}}{Z}|01\rangle\nonumber \\
 &&=\xi_{+, +}(\mathcal{T})\Big[-i\kappa Q(\mathcal{T})\big(P(t)P(\tau)+\kappa^2 Q(\tau)Q(t)\big)\langle 01| \tilde{\rho}_s(t)|01\rangle
 +\kappa^2 Q(\mathcal{T})\big(Q(t)P^*(\tau)-P(t)Q(\tau)\big)\langle 10| \tilde{\rho}_s(t)|01\rangle\Big]\nonumber \\
 &&~+\xi_{+, -}(\mathcal{T}) \Big[i\kappa(Q(t)P^*(\mathcal{T})P(\tau) -Q(\tau)P(t)P(\mathcal{T}))\langle 01| \tilde{\rho}_s(t)|01\rangle
+ \big(\kappa^2 Q(t)Q(\tau)P^*(\mathcal{T})+P(t)P(\mathcal{T})P^*(\tau)\big)\langle 10| \tilde{\rho}_s(t)|01\rangle\Big].\nonumber \\
\label{mat1}
\end{eqnarray}
The evolution of the system parts of the right-hand side of Eq. (\ref{1st_mntxt_s}) can be calculated using the following relations:
\begin{eqnarray}
 e^{-i H_s^{L}t}|10\rangle&=&[P(t)^* |10\rangle -i\kappa Q(t) |01\rangle]e^{i\frac{J_{\parallel}}{4}t} ,
\label{basis1}
\end{eqnarray}
and
\begin{eqnarray}
 e^{-i H_s^{L}t}|01\rangle&=&[P(t) |01\rangle -i\kappa Q(t) |10\rangle]e^{i\frac{J_{\parallel}}{4}t},
 \label{basis2}
\end{eqnarray}
with $P(t)=\rm {cos}\Big (t \sqrt{\frac{\Delta \varepsilon^2}{4}+\kappa^2} \Big)+i\frac{\Delta \varepsilon}{2}
 \frac{\rm{sin}\Big ( t \sqrt{\frac{\Delta \varepsilon^2}{4}+\kappa^2} \Big)}{\sqrt{\frac{\Delta \varepsilon^2}{4}+\kappa^2}}$ and 
 $Q(t)=\frac{\rm{sin}\Big (t \sqrt{\frac{\Delta \varepsilon^2}{4}+\kappa^2} \Big)}{\sqrt{ \frac{\Delta \varepsilon^2}{4}+\kappa^2}}$.
For $\kappa\ll\Delta\varepsilon$, we can approximate Eq. (\ref{mat1}) as
\begin{eqnarray}
 \langle 10|\sum_{n}~_{ph}\langle n |
 \tilde{H}_I^L(t)\tilde{H}_I^L(\tau)| n \rangle_{ph} \tilde{\rho}_s(t) \frac{e^{-\beta \bar{\omega}_{n}}}{Z}|01\rangle&=&\xi_{+, -}(\mathcal{T}) P(t)P(\mathcal{T})P^*(\tau)
 \langle 10| \tilde{\rho}_s(t)|01\rangle\nonumber \\
 &=&\xi_{+, -}(\mathcal{T})e^{i\Delta\varepsilon\mathcal{T}}\langle 10| \tilde{\rho}_s(t)|01\rangle.
 \label{approx1}
\end{eqnarray}
The second term is given by
\begin{eqnarray}
 &&\langle 10|\sum_{n}~_{ph}\langle n |
 \tilde{H}_I^L(t)\tilde{\rho}_s(t)\otimes R_{0}\tilde{H}_I^L(\tau)| n \rangle_{ph}|01\rangle\nonumber \\
 &&=\xi_{+, +}(-\mathcal{T})\Big[\langle 10| \tilde{\rho}_s(t)|10\rangle \big( -i\kappa P(t)P^2(\tau)Q(t)+i\kappa^3 Q(t)Q^2(\tau)P^*(t)\big)\nonumber \\
 &&~~~~~~~+\langle 10| \tilde{\rho}_s(t)|01\rangle \big( \kappa^2 P(t)P(\tau)Q(t)Q(\tau)+\kappa^2 Q(t)Q(\tau)P^*(t)P^*(\tau)\big)\nonumber \\ 
 &&~~~~~~~+\langle 01| \tilde{\rho}_s(t)|10\rangle \big( P^2(t)P^2(\tau)+\kappa^4 Q^2(t)Q^2(\tau)\big)\nonumber \\
 &&~~~~~~~+\langle 01| \tilde{\rho}_s(t)|01\rangle \big(i\kappa P^2(t)P(\tau)Q(\tau)-i\kappa^3 Q^2(t)Q(\tau)P^*(\tau)\big)\Big]\nonumber \\
 &&~~~+\xi_{+, -}(-\mathcal{T})\Big[ \langle 10| \tilde{\rho}_s(t)|10\rangle \big(-i\kappa^3 P(t)Q(t)Q^2(\tau)+i\kappa Q(t)P^*(t)P^2(\tau)\big)\nonumber \\
 &&~~~~~~~+\langle 10| \tilde{\rho}_s(t)|01\rangle\big(-\kappa^2 P(t)P^*(\tau)Q(t)Q(\tau)-\kappa^2 Q(t)P^*(t)Q(\tau)P(\tau)\big)\nonumber\\
 &&~~~~~~~+\langle 01| \tilde{\rho}_s(t)|10\rangle\big(\kappa^2 P^2(t)Q^2(\tau)+\kappa^2 Q^2(t)P^2(\tau)\big)\nonumber \\
 &&~~~~~~~+\langle 01| \tilde{\rho}_s(t)|01\rangle\big(i\kappa^3 Q^2(t)Q(\tau)P(\tau)-i\kappa P^2(t)P^*(\tau)Q(\tau)\big)\Big].
 \label{mat2}
\end{eqnarray}
For $\kappa\ll\Delta\varepsilon$, we get 
\begin{eqnarray}
 \langle 10|\sum_{n}~_{ph}\langle n |
 \tilde{H}_I^L(\tau)\tilde{\rho}_s(t)\otimes R_{0}\tilde{H}_I^L(t)| n \rangle_{ph}|01\rangle&=&
 \xi_{+, +}(-\mathcal{T})\langle 01| \tilde{\rho}_s(t)|10\rangle P^2(t)P^2(\tau)\nonumber \\
 &=& \xi_{+, +}(-\mathcal{T})\langle 01| \tilde{\rho}_s(t)|10\rangle e^{i\Delta\varepsilon(t+\tau)}.
 \label{approx2}
\end{eqnarray}
The third term is expressed as
\begin{eqnarray}
 &&\langle 10|\sum_{n}~_{ph}\langle n |
 \tilde{H}_I^L(\tau)\tilde{\rho}_s(t)\otimes R_{0}\tilde{H}_I^L(t)| n \rangle_{ph}|01\rangle\nonumber \\
 &&=\xi_{+, +}(\mathcal{T})\Big[\langle 10| \tilde{\rho}_s(t)|10\rangle \big( -i\kappa P(\tau)P^2(t)Q(\tau)+i\kappa^3 Q(\tau)Q^2(t)P^*(\tau)\big)\nonumber \\
 &&~~~~~~~+\langle 10| \tilde{\rho}_s(t)|01\rangle \big( \kappa^2 P(\tau)P(t)Q(t)Q(\tau)+\kappa^2 Q(\tau)Q(t)P^*(\tau)P^*(t)\big)\nonumber \\ 
 &&~~~~~~~+\langle 01| \tilde{\rho}_s(t)|10\rangle \big( P^2(\tau)P^2(t)+\kappa^4 Q^2(\tau)Q^2(t)\big)\nonumber \\
 &&~~~~~~~+\langle 01| \tilde{\rho}_s(t)|01\rangle \big(i\kappa P^2(\tau)P(t)Q(t)-i\kappa^3 Q^2(\tau)Q(t)P^*(t)\big)\Big]\nonumber \\
 &&~~~+\xi_{+, -}(\mathcal{T})\Big[ \langle 10| \tilde{\rho}_s(t)|10\rangle \big(-i\kappa^3 P(\tau)Q(\tau)Q^2(t)+i\kappa Q(\tau)P^*(\tau)P^2(t)\big)\nonumber \\
 &&~~~~~~~+\langle 10| \tilde{\rho}_s(t)|01\rangle\big(-\kappa^2 P(\tau)P^*(t)Q(\tau)Q(t)-\kappa^2 Q(\tau)P^*(\tau)Q(t)P(t)\big)\nonumber\\
 &&~~~~~~~+\langle 01| \tilde{\rho}_s(t)|10\rangle\big(\kappa^2 P^2(\tau)Q^2(t)+\kappa^2 Q^2(\tau)P^2(t)\big)\nonumber \\
 &&~~~~~~~+\langle 01| \tilde{\rho}_s(t)|01\rangle\big(i\kappa^3 Q^2(\tau)Q(t)P(t)-i\kappa P^2(\tau)P^*(t)Q(t)\big)\Big].
 \label{mat3}
\end{eqnarray}
For $\kappa\ll\Delta\varepsilon$, we obtain
\begin{eqnarray}
 \langle 10|\sum_{n}~_{ph}\langle n |
 \tilde{H}_I^L(\tau)\tilde{\rho}_s(t)\otimes R_{0}\tilde{H}_I^L(t)| n \rangle_{ph}|01\rangle&=&
 \xi_{+, +}(\mathcal{T})\langle 01| \tilde{\rho}_s(t)|10\rangle P^2(\tau)P^2(t)\nonumber \\
 &=& \xi_{+, +}(\mathcal{T})\langle 01| \tilde{\rho}_s(t)|10\rangle e^{i\Delta\varepsilon(t+\tau)}.
 \label{approx3}
\end{eqnarray}

Lastly, the fourth term reads
\begin{eqnarray}
 &&\langle 10|\sum_{n}~\tilde{\rho}_s(t) \frac{e^{-\beta \bar{\omega}_{n}}}{Z}_{ph}\langle n |
 \tilde{H}_I^L(\tau)\tilde{H}_I^L(t)| n \rangle_{ph} |01\rangle\nonumber \\
 &&=\xi_{+, +}(-\mathcal{T})\Big[i\kappa Q(\mathcal{T})\big(P(t)P(\tau)+\kappa^2 Q(\tau)Q(t)\big)\langle 10| \tilde{\rho}_s(t)|10\rangle
 +\kappa^2 Q(\mathcal{T})\big(Q(t)P^*(\tau)-P(t)Q(\tau)\big)\langle 10| \tilde{\rho}_s(t)|01\rangle\Big]\nonumber \\
 &&~+\xi_{+, -}(-\mathcal{T}) \Big[i\kappa(-Q(t)P^*(\mathcal{T})P(\tau) +Q(\tau)P(t)P(\mathcal{T}))\langle 10| \tilde{\rho}_s(t)|10\rangle
+ \big(\kappa^2 Q(t)Q(\tau)P^*(\mathcal{T})+P(t)P(\mathcal{T})P^*(\tau)\big)\langle 10| \tilde{\rho}_s(t)|01\rangle\Big].\nonumber \\
\label{mat4}
\end{eqnarray}
For $\kappa\ll\Delta\varepsilon$, we can approximate the above expression as
\begin{eqnarray}
 \langle 10|\sum_{n}~\tilde{\rho}_s(t) \frac{e^{-\beta \bar{\omega}_{n}}}{Z}_{ph}\langle n |
 \tilde{H}_I^L(\tau)\tilde{H}_I^L(t)| n \rangle_{ph} |01\rangle&=&\xi_{+, -}(-\mathcal{T}) P(t)P(\mathcal{T})P^*(\tau)
 \langle 10| \tilde{\rho}_s(t)|01\rangle\nonumber \\
 &=&\xi_{+, -}(-\mathcal{T})e^{i\Delta\varepsilon\mathcal{T}}\langle 10| \tilde{\rho}_s(t)|01\rangle.
 \label{approx4}
\end{eqnarray}
Putting Eqs. (\ref{approx1}), (\ref{approx2}), (\ref{approx3}), and (\ref{approx4}) in Eq. (5) of the main text 
and using the equalities in Eq. (\ref{matcon}), one obtains  Eq. (8) in the main text. In a similar way, one can deduce Eq. (9) for the diagonal element. 
Eqs. (6) and (7) can be deduced by using Eqs. (\ref{mat1}), (\ref{mat2}), (\ref{mat3}), and (\ref{mat4}) without the approximation 
$\kappa\ll\Delta\varepsilon$ and re-expressing the terms in \{$|\epsilon_s\rangle$, $|\epsilon_t\rangle$\} basis.
It should be noted that, at zero detuning,
the only system excitation is given by $\delta \epsilon=2\kappa=J^{\rm mf}_{\perp}$.
\end{widetext}
\end{document}